\documentstyle[12pt]{article}

\catcode`\@=11
\long\def\@makefntext#1{
\protect\noindent \hbox to 3.2pt {\hskip-.9pt
$^{{\ninerm\@thefnmark}}$\hfil}#1\hfill}                

\def\thefootnote{\fnsymbol{footnote}}
 \def\@makefnmark{\hbox to 0pt{$^{\@thefnmark}$\hss}}  

\def\ps@myheadings{\let\@mkboth\@gobbletwo
\def\@oddhead{\hbox{}
\rightmark\hfil\ninerm\thepage}
\def\@oddfoot{}\def\@evenhead{\ninerm\thepage\hfil
\leftmark\hbox{}}\def\@evenfoot{}
\def\sectionmark##1{}\def\subsectionmark##1{}}

\renewcommand{\thefootnote}{\fnsymbol{footnote}}

\newcounter{sectionc}\newcounter{subsectionc}\newcounter{subsubsectionc}
\renewcommand{\section}[1] {\vspace{0.6cm}\addtocounter{sectionc}{1}
\setcounter{subsectionc}{0}\setcounter{subsubsectionc}{0}\noindent
        {\bf\thesectionc. #1}\par\vspace{0.4cm}}
\renewcommand{\subsection}[1] {\vspace{0.6cm}\addtocounter{subsectionc}{1}
        \setcounter{subsubsectionc}{0}\noindent
        {\it\thesectionc.\thesubsectionc. #1}\par\vspace{0.4cm}}
\renewcommand{\subsubsection}[1]
{\vspace{0.6cm}\addtocounter{subsubsectionc}{1}
        \noindent {\rm\thesectionc.\thesubsectionc.\thesubsubsectionc.
        #1}\par\vspace{0.4cm}}

\newcounter{appendixc}
\newcounter{subappendixc}[appendixc]
\newcounter{subsubappendixc}[subappendixc]

\renewcommand{\appendix}[1] {\vspace{0.6cm}
        \refstepcounter{appendixc}
        \setcounter{figure}{0}
        \setcounter{table}{0}
        \setcounter{equation}{0}
        \renewcommand{\thefigure}{\Alph{appendixc}.\arabic{figure}}
        \renewcommand{\thetable}{\Alph{appendixc}.\arabic{table}}
        \renewcommand{\theappendixc}{\Alph{appendixc}}
        \renewcommand{\theequation}{\Alph{appendixc}.\arabic{equation}}
        \noindent{\bf Appendix \theappendixc #1}\par\vspace{0.4cm}}

\def\abstracts#1{{
        \centering{\begin{minipage}{30pc}\tenrm\baselineskip=12pt\noindent
        \centerline{\tenrm ABSTRACT}\vspace{0.3cm}
        \parindent=0pt #1
        \end{minipage}}\par}}

\renewenvironment{thebibliography}[1]
        {\begin{list}{\arabic{enumi}.}
        {\usecounter{enumi}\setlength{\parsep}{0pt}
\setlength{\leftmargin 1.25cm}{\rightmargin 0pt}
         \setlength{\itemsep}{0pt} \settowidth
        {\labelwidth}{#1.}\sloppy}}{\end{list}}

\newcounter{itemlistc}
\newcounter{romanlistc}
\newcounter{alphlistc}
\newcounter{arabiclistc}

\newcommand{\fcaption}[1]{
        \refstepcounter{figure}
        \setbox\@tempboxa = \hbox{\tenrm Fig.~\thefigure. #1}
        \ifdim \wd\@tempboxa > 6in
           {\begin{center}
        \parbox{6in}{\tenrm\baselineskip=12pt Fig.~\thefigure. #1}
            \end{center}}
        \else
             {\begin{center}
             {\tenrm Fig.~\thefigure. #1}
              \end{center}}
        \fi}

\newcommand{\tcaption}[1]{
        \refstepcounter{table}
        \setbox\@tempboxa = \hbox{\tenrm Table~\thetable. #1}
        \ifdim \wd\@tempboxa > 6in
           {\begin{center}
        \parbox{6in}{\tenrm\baselineskip=12pt Table~\thetable. #1}
            \end{center}}
        \else
             {\begin{center}
             {\tenrm Table~\thetable. #1}
              \end{center}}
        \fi}

\def\@citex[#1]#2{\if@filesw\immediate\write\@auxout
        {\string\citation{#2}}\fi
\def\@citea{}\@cite{\@for\@citeb:=#2\do
        {\@citea\def\@citea{,}\@ifundefined
        {b@\@citeb}{{\bf ?}\@warning
        {Citation `\@citeb' on page \thepage \space undefined}}
        {\csname b@\@citeb\endcsname}}}{#1}}

\newif\if@cghi
\def\cite{\@cghitrue\@ifnextchar [{\@tempswatrue
        \@citex}{\@tempswafalse\@citex[]}}
\def\citelow{\@cghifalse\@ifnextchar [{\@tempswatrue
        \@citex}{\@tempswafalse\@citex[]}}
\def\@cite#1#2{{$\null^{#1}$\if@tempswa\typeout
        {IJCGA warning: optional citation argument
        ignored: `#2'} \fi}}

\def\fnt#1#2{\footnotetext{\kern-.3em
        {$^{\mbox{\sevenrm #1}}$}{#2}}}

 1
 1
 1

\font\tenbf=cmbx10
\font\tenrm=cmr10
\font\tenit=cmti10

\font\ninerm=cmr9

\begin{document}

\newcommand{\be}{\begin{equation}}
\newcommand{\ee}{\end{equation}}
\newcommand{\bea}{\begin{eqnarray}}
\newcommand{\eea}{\end{eqnarray}}
\newcommand{\len}{\lefteqn}
\newcommand{\nn}{\nonumber}
\newcommand{\muh}{\hat\mu}
\newcommand{\dlr}{\stackrel{\leftrightarrow}{D} _\mu}
\newcommand{\vnew}{$V^{\rm{NEW}}$}
\newcommand{\vecp}{$\vec p$}
\newcommand{\dof}{{\rm d.o.f.}}
\newcommand{\prd}{Phys. Rev. \underline}
\newcommand{\pl}{Phys. Lett. \underline}
\newcommand{\prl}{Phys. Rev. Lett. \underline}
\newcommand{\np}{Nucl.Phys. \underline}
\newcommand{\vvp}{v_B\cdot v_D}
\newcommand{\dl}{\stackrel{\leftarrow}{D}}
\newcommand{\dr}{\stackrel{\rightarrow}{D}}
\newcommand{\mev}{{\rm MeV}}
\newcommand{\gev}{{\rm GeV}}
\newcommand{\calp}{{\cal P}}
\newcommand{\pinc}{\vec p \hskip 0.3em ^{inc}}
\newcommand{\pout}{\vec p \hskip 0.3em ^{out}}
\newcommand{\ptr}{\vec p \hskip 0.3em ^{tr}}
\newcommand{\pbr}{\vec p \hskip 0.3em ^{br}}
\newcommand{\no}{\noindent}
\newcommand{\ra}{\rightarrow}
\def\glgt{$\Gamma_L/\Gamma_{tot}$ }
\def\gs {$\Gamma(B\to K^\ast \psi)/\Gamma(B\to K \psi)$ }
\def\und{\underline}
\def\beq{\begin{equation}}
\def\eeq{\end{equation}}
\def\noi{\noindent}
\def\bea{\begin{eqnarray}}
\def\eea{\end{eqnarray}}
\def\bei{\begin{itemize}}
\def\eei{\end{itemize}}
\def\bpsi{$B\ra K^{(\ast)} \psi$ }
\def\dk{$D\ra K^{(\ast)}l \nu$ }
\def\hh {heavy-to-heavy }
\def\hhs {heavy-to-heavy scaling }
\def\hl {heavy-to-light }
\def\hls {heavy-to-light scaling }
\def\qq {$q^2$ }
\def\qmax {$q^2_{max}$ }
\def\qqz {$q^2=0$ }
\renewcommand{\thefootnote}{\fnsymbol{footnote}}
\def\fourthrm{\font\fourthrm=cmr12 at 14pt}
\def\ved{\vec q\, ^2}
\begin{flushright} LPTHE Orsay-94/105\\ hep-ph/9412229
\end{flushright}
\centerline{\tenbf THE $B \to K^\ast \psi$  POLARIZATION PUZZLE}
\baselineskip=16pt
\vspace{0.8cm}
\centerline{\tenrm R. Aleksan,}
\baselineskip=13pt
\centerline{\tenit Centre d'Etudes Nucl\'eaires de Saclay, DPhPE, 91191
Gif-sur-Yvette, France}
\vspace{0.3cm}
\centerline{\tenrm A. Le Yaouanc, L. Oliver, O. P\`ene and J.-C. Raynal}
\baselineskip=13pt
\centerline{\tenit Laboratoire de Physique Th\'eorique et Hautes
Energies\footnote{Laboratoire
associ\'e au
Centre National de la Recherche Scientifique - URA 63}}
\baselineskip=12pt
\centerline{\tenit Universit\'e de Paris XI, B\^atiment 211, 91405 Orsay Cedex,
France}
\vspace{0.3cm}
\centerline{Contribution to the First Arctic Workshop on Future Physics
 and Accelarators,}
\centerline{ Saariselk\"a, Finland, August 20-26 1994.}
\vspace{0.3cm}
\centerline{\tenrm Presented by}
\vspace{0.3cm}
\centerline{\tenrm O. P\`ene}

\vspace{0.9cm}
\abstracts{We point out that current estimates of form factors fail to explain
the
non-leptonic decays $B \to \psi K(K^{\ast})$ and that the combination of data
on the semi-leptonic decays $D \to K(K^{\ast})\ell \nu$ and
on the non-leptonic decays $B \to \psi K(K^{\ast})$ (in particular recent
po\-la\-ri\-za\-tion data)
severely constrain the form (normalization and $q^2$ dependence) of the
heavy-to-light meson form
factors, if we assume the factorization hypothesis for the latter. From a
simultaneous fit to \bpsi and \dk data we find that
strict heavy quark limit scaling laws applied to the form factors
do not hold  when going from $D$ to $B$
and must have large
corrections that make softer the dependence on the masses.
This is in contrast with the matrix elements themselves which
are found to need smaller $1/m_Q$ corrections to the asymptotic
heavy quark scaling laws.
 We also find that
$A_1(q^2)$ should increase slower with \qq than $A_2, V, f_+$.
 We propose a simple parametrization of these corrections based on a quark
model or on an extension of the \hhs laws to the \hl case, complemented with an
approximately constant $A_1(q^2)$. This model may be viewed as
 assuming a precocious validity of
strict heavy quark scaling laws for the current matrix elements.
 Although this model reproduces qualitatively the
wanted  features for mass and $q^2$ dependence, and thus reduces the
discrepancy with data,
it is insufficient to reach a full agreement with the experimental
polarization.
In our opinion the puzzle is still there.
}

\vfil
\rm\baselineskip=14pt
\section{Introduction.}

 We heard\cite{oddone} a nice talk about the SLAC B-factory
and the BaBar detector which are planned mainly to detect CP violation
in $B$ meson decays.

We will not repeat why the search for CP violation is of crucial  theoretical
interest in a period where only one type of laboratory experiment,
namely $K_L\to \pi \pi$ has positively found CP violation\footnote{Let us leave
aside
 the question of baryogenesis which does not seem to fit easily in the Standard
Model.}
. The $B$ system presents
 this peculiarity that CP violation is commonly believed to lie within
experimental
reach, in the Standard Model, while non-standard surprises may also happen.
This has motivated
the building of B-factories.

The so-called angle $\beta$ of the unitarity triangle should be measurable
through the $B\to \psi K_S$ CP asymmetries and/or
through the $B\to D^{(\ast)}\overline D^{(\ast)}$ ones\cite{alek18}. The angle
$\alpha$
will not be so easy to measure, and the angle $\gamma$ is still a challenge.
Adding
several channels as in\cite{alek18} increases the stastitics provided that the
relative signs of the different channel are such as not to  wash out the
asymmetry. In the case
of  $B\to D^{(\ast)}\overline D^{(\ast)}$ decays, the Heavy Quark Symmetry
(HQS) helps to
establish these relative signs which turn out not to dilute the CP asymmetry.
In
 ref.\cite{alek03} we considered applying the same trick to
 $B_s\to K^{(\ast)} D_s^{(\ast)}$ decay channels. In this case HQS is not so
helpful.

We used the factorization assumption and could argue about the signs of the
leptonic
and semi-leptonic form-factors. However, one crucial expression comes in, of
the form
$A_1 - c A_2$ where $A_{1,2}$ are semileptonic form factors and $c$ is a
kinematical factor
whose precise value does not matter here. Using some models, or some arguments
from the
HQS applied to heavy-to-light decays, we found that the sign of this expression
was
not easy to settle. At this point we realized that the same expression
$A_1 - c A_2$
with a slightly different kinematical factor $c$ also appeared in the
expression for
the polarization in $B\to \psi K^\ast$, a quantity on which several recent and
rather
precise experimental data exist.

We thus started to look closer  at $B\to \psi K^{(\ast)}$ decays and we
encountered
{\it quite a surprise}. It became obvious to us\cite{beauty-94}$^,$\cite{nous}
that the most popular models
which are commonly used in heavy flavor decays as well as the simple-minded
application
of HQS extrapolated from $D$ semi-leptonic decays severely fail to explain
$B\to K^\ast \psi$ polarization and the ratio $\Gamma(B\to \psi
K^\ast)/\Gamma(B\to \psi K)$.
This conclusion was independently reached by Gourdin, Keum and
Pham\cite{gourdin}.

Such a general failure casts some doubt on the predictions one can extract from
the same models and
simple-minded ideas whenever some precision is wanted as it was the case in our
above-mentioned example
 of the sign of $A_1 - c A_2$ and the consecutive non dilution
of CP asymetries. As long as the puzzle will not be solved, it will at least
demand the
 severest care and the systematic use of $B\to \psi K^\ast$ as a touchstone. We
tried to
stick to such an attitude in\cite{alek03} and we believe we could safely
deliver a positive answer about
the non-dilution of the CP asymmetry.

We can also take   $B\to \psi K^{(\ast)}$ data as a precious source of
phenomenological information
in the domain of $B$ decays where precise data are still missing.
In the following we will first try to convince you that there is really a
problem. We discuss next
what positive knowledge can be extracted from these polarization data. Of
course, this needs
some assumption: we will stick to the factorization assumption.

$B\to \psi K^\ast$ polarization measurements turn out to be a very efficient
touchstone
of our present understanding of the non-leptonic decays of $B$ mesons.

\section{$B\to \psi K^{(\ast)}$ data are hardly compatible with current
estimates.}
\label{sec-bpsi}

	To be definite, let us write the form factors :

\[< P_f|V_{\mu}|P_i> = \left ( p_{\mu}^f + p_{\mu}^j - {m_i^2 - m_f^2 \over
q^2}  q_{\mu} \right
) f_+(q^2) + {m^2_i - m^2_f \over q^2}  q_{\mu} f_0(q^2),\]
\[< V_f|A_{\mu}|P_i > = \left ( m_f + m_i \right ) A_1(q^2)
\left ( \varepsilon_{\mu}^{\ast} - {\varepsilon^{\ast}.q \over q^2} q_{\mu}
\right )
\]
\[- A_2(q^2) {\varepsilon^{\ast}.q \over m_f + m_i}
\left ( p^i_{\mu} + p^f_{\mu} -
{m^2_i - m^2_f \over q^2}  q_{\mu} \right ) + 2 m_f
\ A_0(q^2)  {\varepsilon^{\ast} . q \over q^2} q_{\mu},\]
\beq
< V_f|V_{\mu}|P_i > = i {2 \ V(q^2) \over m_f + m_i}  \varepsilon_{\mu \nu \rho
\sigma} p_i^{\nu} \  p_f^{\rho} \varepsilon^{\ast \sigma}.\label{definite}
\eeq
where we use the convention $\epsilon^{0123}=1$.

Using factorization\cite{svz}$^,$\cite{bsw1}$^,$\cite{bsw2},
one obtains the following amplitudes in
the B meson rest frame:

\beq
A \left ( \bar{B}_d^0 \to \psi K \right ) = - {G \over \sqrt{2}} V_{cb}
V_{cs}^{\ast} \ 2 \ f_{\psi} \
m_B \ f_+(m^2_{\psi} ) a_2p,
 \eeq

\[A^{pv} \left (\bar{B}^O_d \to \psi (\lambda = 0) K^{\ast}(\lambda = 0) \right
) = - {G
\over \sqrt{2}}  V_{cb} V_{cs}^{\ast} m_{\psi} f_{\psi},\]
\beq
\left [ \left ( m_B + m_{K^{\ast}} \right ) \left (
{p^2 + E_{K^{\ast}} E_{\psi} \over m_{K^{\ast}} m_{\psi}} \right )
A_1(m^2_{\psi}) - {m^2_B \over m_B + m_{K^{\ast}}}
{2p^2 \over m_{K^{\ast}} m_{\psi}} A_2(m^2_{\psi}) \right ] a_2,\label{long}
\eeq
\beq
A^{pv} \left ( \bar{B}^0_d \to \psi(\lambda = \pm 1) K^{\ast}(\lambda =
\pm 1) \right ) = - {G \over \sqrt{2}} V_{cb}V_{cs}^{\ast} m_{\psi}
f_{\psi} \left ( m_B + m_{K^{\ast}} \right ) A_1(m^2_{\psi}) a_2,
\eeq
\beq
A^{pc} \left ( \bar{B}^0_d \to \psi(\lambda = \pm 1) K^{\ast}(\lambda =
\pm 1) \right ) = \pm {G \over\sqrt{2}}
V_{cb}V_{cs}^{\ast} m_{\psi} f_{\psi}  {m_B \over m_B + m_{K^{\ast}}} 2
V(m^2_{\psi}) a_2 p.
\eeq

These amplitude are all proportional to $a_2$, i.e. they belong to the
so-called
class II decays.
We see that the non-leptonic data plus the factorization hypothesis can give us
information on the form factors at a different kinematic point ($q^2 =
m^2_{\psi})$ than the data on
semi-leptonic $D$ decays (small $q^2$) or the heavy quark limit QCD scaling
laws (close to
$q^2_{max}$).

The data for the total rates\cite{cleo-94} are:

\[
BR \left ( \bar{B}_d^0 \to \psi K^0 \right ) = (7.5 \pm 2.4 \pm 0.8) \times
10^{-4}
\]

\[
BR \left ( B_d^0 \to \psi K^{\ast 0} \right ) = (16.9 \pm 3.1 \pm 1.8) \times
10^{-4}
\]
\[
BR \left ( {B}^- \to \psi K^- \right ) = (11.0 \pm 1.5 \pm 0.9) \times
10^{-4}
\]
\[
BR \left ( B^- \to \psi K^{\ast -} \right ) = (17.8 \pm 5.1 \pm 2.3) \times
10^{-4}
\]
\noi and  the recent results of ARGUS\cite{cleo2}, CLEO\cite{cleo2} and
CDF\cite{cdf} concerning the $K^\ast$ polarization in the
$\bar{B}_d \to \psi K^{\ast 0}$ decay, are:

\bea
\Gamma_L /\Gamma_{tot} & > & 0.78 \ (95 \  \% \ C.L.)		\quad \mbox{ARGUS}\,
\nonumber \\
\Gamma_L /\Gamma_{tot} & = & 0.80 \pm 0.08 \pm 0.05 	\quad \mbox{CLEO}
\nonumber \\
\Gamma_L /\Gamma_{tot} & = & 0.66 \pm 0.10 ^{+0.10}_{-0.08} 	\quad \mbox{CDF}
\label{rlexp}\eea

\noindent where $\Gamma_L$ is the partial width for the longitudinal
polarization whose amplitude is
given by (\ref{long}). \par

	As already pointed out, these decays are affected by
the phenomenological factor $a_2$ which is not well known.
To avoid this uncertainty, we will consider the
ratio of the total rates

\beq
R \equiv {\Gamma \left ( \bar{B}_d^0 \to \psi K^{\ast 0} \right ) \over \Gamma
\left (
\bar{B}_d^0 \to \psi K^{0} \right )} = 1.64 \pm 0.34 \quad\mbox{CLEO
II}^{13},\,\,
 \label{rstar}\eeq

\noi  and the polarization ratio for $\psi K^{\ast 0}$~:

\beq
R_L \equiv {\Gamma_L \left ( \bar{B}_d^0 \to \psi K^{\ast 0} \right ) \over
\Gamma_{tot} \left ( \bar{B}_d^0 \to \psi K^{\ast 0} \right ),}
\label{rl}\eeq
which are independent of $a_2$.

{}From these formulae one can already conclude qualitatively that :

i) To get $R_L$ sufficiently large, one needs $V/A_1$ and $A_2/A_1$ to be small
enough.

ii) To get $R$ not too large $f_+/A_1$ must not be too small.

 We will consider the predictions for these ratios from the following
theoretical schemes :
\bei
\item	{1.} Pole model of Bauer, Stech and Wirbel (BSWI)\cite{bsw1}.
\item	{2.} Pole-dipole model of Neubert et al. (BSWII)\cite{bsw2}.
\item	{3.} Quark model of Isgur, Scora, Grinstein and Wise (ISGW)\cite{isgw}.
\item	{4.} QCD sum rules (QCDSR)\cite{ball}.
\item	{5.} Lattice QCD\cite{abada}.
\eei
	The results are given in Table \ref{tab-RRL}.  The
conclusion is that there is a problem for all known theoretical
schemes since both ratios $R$ and $R_L$ cannot be described at the same time. A
priori there are three possible explanations :

i) The theoretical schemes for {\it form factors} are to be blamed for the
failure.

ii) The experimental numbers are not to be trusted too  much.

iii) The basic BSW factorization assumption, which allows to relate \bpsi to
the form factors, is wrong  for class II decays.

In section 3, we will explore the first possibility
on general grounds using the data on \dk and \bpsi combined through HQS.
To make the study more quantitative, we will propose a reasonable Ansatz which
 presents in our opinion the general features that are favored by phenomenology
and by some theoretical considerations. This Ansatz, although meant
to reconcile both sets of data without violating HQS,
still leaves a 2-3 $\sigma$ discrepancy between \dk and \bpsi data.

Besides the experimental failure apparent in table 1, the quoted popular
quark models also present theoretical problems\cite{nous}: the BSW models do
not
satisfy the heavy quark scaling laws when the mass goes to infinity,
while the ISGW satifies only the heavy-to-light scaling law when the
initial mass goes to infinity, failing to satisfy the heavy-to-heavy scaling
laws when
both the initial and final masses go to infinity. We will not elaborate further
on this issue in this talk.

\begin{table}
\centering
\begin{tabular} {|c|c|c|c|c|c|}
\hline
 &  $\frac{\Gamma(K^\ast)}{\Gamma(K)}$ & $\frac{\Gamma_L}{\Gamma_{tot}}$ &
$\frac {A_2^{sb}(m_\psi^2)}{A_1^{sb}(m_\psi^2)}$&$
 \frac {V^{sb}(m_\psi^2)}{A_1^{sb}(m_\psi^2)}$&$ \frac
{f_+^{sb}(m_\psi^2)}{A_1^{sb}(m_\psi^2)}$\\ \hline
BSWI\cite{bsw1}&4.23	&0.57 & 1.01& 1.20 &1.23\\ \hline
BSWII\cite{bsw2}&1.61	&0.36 &1.41  & 1.77 &1.82\\ \hline
ISGW\cite{isgw}& 1.71&0.07 & 2.00 & 2.58 &2.30\\ \hline
QCDSR\cite{ball}$^,$\cite{ballp}& 7.60&0.36 & 1.19 & 2.66  & 1.77\\ \hline
Lattice (a) & $ 3.5\pm 2.5$ & $0.47\pm 0.11 $& & &\\ \hline
Lattice (b)& $ 1.9\pm 1.4$ & $0.27\pm0.16$ & & &\\ \hline
Our Ansatz & 2.15 & 0.45 & 1.08 & 2.16 & 1.86 \\ \hline
CLEO II\cite{cleo2}$^,$\cite{browder} 	&    $1.64\pm 0.34$   &	    $0.8\pm0.1$
&
& &\\
CDF\cite{cdf}	&    -   &	    $0.66\pm0.1$ & & &\\\hline
\end{tabular}
\caption{\it Comparison of different models, a QCD Sum Rules calculation,
a lattice calculation,
 and our prefered Ansatz (as defined later on) to experiment. The mass and
$q^2$ extrapolation for lattice results are detailed in ref.$^{17}$. Lattice
(a)
uses a kinematical pole in $A_2/A_1$ while Lattice (b) uses a constant
$A_2/A_1$
as $q^2$ varies.}

\label{tab-RRL}
\end{table}

\section{Heavy Quark Symmetry and the relation to $D$ decays.}
\label{sec-class}
\vspace{-0.7cm}
\subsection{Setting  the Problem.}
\label{sec-setting}
\vspace{-0.35cm}
	Our aim was to perform a combined  experimental and theoretical study of
form factors, simultaneously for both of \dk and \bpsi decays, assuming
factorization for the latter. We wanted to proceed as independently
as possible
of the detailed theoretical approaches, using {\it general Ans\"atze
that respect the \hl asymptotic scaling laws}, some of them being complemented
by ideas derived from \hhs law formulae. Only guided by  rigorous theoretical
laws and some commonly admitted theoretical prejudices, we will try to display
general trends suggested by the experiment. However it turns out that
experiment, as
it stays today, is not easy to account for in a theoretically reasonable
manner. We will also advocate the use of a Quark Model inspired Ansatz, eq.
(\ref{sp}),
 an extension of some \hhs relations to the \hl
system. Although not fully successful, this model is able to account roughly
for a large set of data.

	First, let us review available data. Besides  the indirect indications
coming from the above $B \to \psi K(K^{\ast})$ non leptonic data complemented
by
the BSW factorization assumption, there are data on the $D \to K(K^{\ast}) \ell
\nu$ form factors, mainly around $q^2 = 0$. We shall use the world
average\cite{witherell}:

\begin{eqnarray}
f^{sc}_+(0) & = & 0.77 \pm 0.08  \nonumber \\
V^{sc}(0) & = & 1.16 \pm 0.16  \nonumber \\
A^{sc}_1(0) & = & 0.61 \pm 0.05  \qquad  \nonumber \\
A^{sc}_2(0) & = & 0.45 \pm 0.09
\label{dexp}\end{eqnarray}

\bea
V^{sc}(0)/A^{sc}_1(0) = 1.9 \pm 0.25 \nonumber \\
A^{sc}_2(0)/A^{sc}_1(0) = 0.74 \pm 0.15  \ \ \ .
\label{drexp}\eea

As to the $q^2$ dependence the indications are poor except for the $f_+$ form
factor, where good indications seem to support the relevant vector meson pole
dominance. We have used\cite{nous} these indications for the \qq dependence to
advocate a pole-like \qq dependence of $f_+, A_2$ and $V$, and a flat $A_1$.
However, in this talk we will concentrate on the evolution from $D$ to $B$ of
the {\it
ratios} between different form factors ($A_2/A_1$, $V/A_1$, etc.) for which we
can formulate more general statements.
 The advantage of discussing first the ratios is that we
can draw more direct conclusions from the  data
before considering absolute branching ratios which also involve
the unknown $a_2$ parameter.

\subsection{Asymptotic Scaling Laws for the Heavy-to-light Form Factors.}
\label{sec-asym}

	What can be learned from the theory ? The only exact results take the form of
asymptotic theorems\cite{isgurhl} valid for the initial quark mass $m_Q$
large with respect to the typical scale of QCD, $\Lambda$,  to the
final meson mass, $m_f$, and to the final momentum,
 $\vec q$\footnote{Unless specified otherwise, we use the initial
meson rest frame.}:

\[ \frac{\hat f_+(\ved)}{c_+},\quad \frac{\hat V(\ved)}{c_V},\quad
\frac{ \hat A_2(\ved)}{c_2} = {m_Q}^{\frac 1 2}
\left(1+O\left(\frac \Lambda{m_Q}\right)+
O\left(\frac {|\vec q|}{m_Q}\right)+O\left(\frac {m_f}{m_Q}\right)\right)\]

\be \frac{\hat A_1(\ved)}{c_1} = \left({m_Q}\right)^{-\frac 12}
\left(1+O\left(\frac
\Lambda{m_Q}\right)+
O\left(\frac {|\vec q|}{m_Q}\right)+O\left(\frac
{m_f}{m_Q}\right)\right)\label{hl}\ee
where the $c_+, c_2, c_V, c_1$ are unknown constants and
where we have used ``hats'' on form factors to indicate that they depend
on the {\it three-momentum, the natural variable in the \hl case}:
\beq
\hat f(\ved) = f(q^2), \qquad \mbox{with}\qquad \ved = \left(\frac
{m_i^2+m_f^2-q^2}{2 m_i}\right)^2-m_f^2\label{hats}
\eeq

The asymptotic scaling law (\ref{hl}) allows to  relate the
form factors, say $D \to K$ and $B \to K$, at small recoil
$|\vec{q}| \ll m_D$ (i.e. close to  $q^2_{max}$ for each process).
In particular, {\it $A_2/A_1$ scales like $m_Q$}. This has dramatic
consequences on
the $B\to \psi K^\ast$ polarization as we shall now see.

\subsection{Failure of the Simple-minded Extrapolation from $D$ to $B$
given the Asymptotic Scaling Laws.}
\label{sec-failu}
\vskip -0.35 cm

Consider the extrapolation from $D\ra K^{(\ast)} l\nu$ data
at $q^2=0$, according to the
\hl asymptotic scaling law. Since the momentum $\vec q$ is different in the two
above-mentioned sets of data, an hypothesis on the $q^2$ dependence is needed.

The ratio \glgt is given by:

\be \frac{\Gamma_L(B\to K^\ast \psi)}{\Gamma_{tot}(B\to K^\ast \psi)}=\frac
{\left(3.162 -1.306 \frac {A_2^{sb}(m_\psi^2)}{A_1^{sb}(m_\psi^2)}
\right)^2}{2\left[1+0.189\left(\frac{V^{sb}(m_\psi^2)}
{A_1^{sb}(m_\psi^2)}\right)^2
\right] +\left(3.162 -1.306 \frac {A_2^{sb}(m_\psi^2)}{A_1^{sb}(m_\psi^2)}
\right)^2}\label{rlform}\ee
where the indices $sb$ indicate that we deal with the $b\to s$ form factors.

{}From this expression it is apparent that $A_2/A_1$ must not be too
large\footnote{Strictly speaking very large values, $A_2/A_1\ge 3.9$,  could
also
account for a large $R_L$, but these are unrealistic.} in view of the large
experimental value of $R_L$ (\ref{rlexp}), all the more if $V/A_1$ is large.
For example, setting $V=0$ we get the very conservative upper bound $A_2/A_1
\le 1.3$ for $R_L>0.5$. For a more realistic value of $V/A_1\simeq 2$, the
upper bound becomes $A_2/A_1 \le 1$. Now, according to strict application of
the asymptotic scaling laws described above, $A_2/A_1$ ($V/A_1$) would be
multiplied at fixed $\vec q$ by $m_B/m_D= 2.83$. From the central experimental
$D$
value, $A^{sc}_2/A^{sc}_1=0.74$ ($V^{sc}/A^{sc}_1=1.9$), one gets
$A^{sb}_2/A^{sb}_1=2.09$ ($V^{sb}/A^{sb}_1= 5.38$) at $q^2= 16.56$ GeV
(corresponding in $B$ decay to the same $\ved$ as $q^2=0$ in $D$ decay). This
is in
drastic contradiction with experiment unless there is an  unexpectedly strong
$q^2$ variation down to $q^2=m_{\psi}^2$. A naive insertion of these values in
eq. (\ref{rlform}) would indeed give $R_L=0.014$ which is 4 to 5 sigmas away
from the most favorable CDF value. Clearly the message is that {\it a softening
of the increase  with respect to the asymptotic scaling law is required}.

It is now useful to consider the product $R(1-R_L)$ where the definitions of
eqs (\ref{rstar}) and (\ref{rl}) have been used:
\be R(1- R_L) = 2.162\left(\frac
{A_1^{sb}(m_\psi^2)}{f_+^{sb}(m_\psi^2)}\right)^2
\left[1+0.189\left(\frac{V^{sb}(m_\psi^2)}{A_1^{sb}(m_\psi^2)}\right)^2\right]
\label{beau}\ee

This gives a lower bound for $f_+/A_1$. For the conservative upper
bounds $R\le 2.5$ and $1-R_L\le 0.5$,  and setting still more conservatively
$V$ to zero, we get $f_+/A_1 \ge 1.32$. For a more realistic estimate,
$V/A_1\simeq 2$, and $R\le 2.0$ we get $f_+/A_1  \ge 1.9$.
Contrarily to our discussion in the preceeding paragraph,  we find here a lower
bound which in itself is compatible with the hard scaling behaviour but not
with such a soft scaling as required for $A_2/A_1$ (remember we had $A_2/A_1\le
1.3$  for $V=0$ and for a more realistic $V/A_1$, $A_2/A_1\le 1$). {\it Clearly
the trend for $f_+/A_1$ is somewhat opposite to the one for $A_2/A_1$ }.
To solve this problem, Cheng and Tseng\cite{cheng} have assumed a monopole form
for
$f_+$ and $A_1$ and a dipole form for $A_2$ and $V$. This implies a pole
behavior for
$A_2/A_1$ leading to a reduced $A^{sb}_2/A^{sb}_1$ and a constant $f_+/A_1$,
thus keeping it large enough.

\subsection{Reminder about Asymptotic Scaling Laws for Heavy-to-heavy
Transitions.}

It is well known that a much stronger set of relations than the one in
subsection 3.2 comes from the Isgur-Wise scaling laws\cite{isgur}
for transition form factors between two heavy quarks. Using the notations
in\cite{nr} :

\[{ \sqrt{4 m_{P_i} m_{P_f}} \over  m_{P_i} + m_{P_f} }  f_+(q^2) =
{ \sqrt{4 m_{P_i} m_{P_f}} \over  m_{P_i} + m_{P_f} } {f_0(q^2) \over  1 -
{q^2 \over \left ( m_{P_i} + m_{P_f} \right )^2}} =
{ \sqrt{4 m_{P_i} m_{V_f}} \over m_{P_i} + m_{V_f} } V(q^2) =\]

\beq
= { \sqrt{4 m_{P_i} m_{V_f}} \over m_{P_i} + m_{V_f} } A_0(q^2) = { \sqrt{4
m_{P_i}
m_{V_f}} \over  m_{P_i} + m_{V_f} } A_2(q^2) = { \sqrt{4 m_{P_i} m_{V_f}}
\over m_{P_i} + m_{V_f} }{A_1(q^2) \over  1 - {q^2 \over \left ( m_{P_i} +
m_{V_f} \right )^2} }  = \xi (v_i.v_f)
\label{hh}\eeq
for $m_{P_i}$,  $m_{P_f}$ and $m_{V_f}$ much larger than the typical QCD scale,
$\Lambda$. In the same limit $m_{P_f}$ and
$m_{V_f}$ are equal, and our writing of  different masses is only
meant for later use in the real subasymptotic regime, where they are very
different ($m_K\ne m_{K^\ast}$).

The denominator that divides $A_1(q^2)$ is a straightforward consequence of the
heavy quark symmetry and of the definition of the different form factors. It
has not the meaning of a dynamical pole related to some intermediate
state. It is still in the mathematical sense a pole of the ratio
$A_2(q^2)/A_1(q^2)$ etc, and we shall call it for simplicity the ``kinematical
pole''.

In the domain of mass with $\Lambda\ll
m_f\ll m_{P_i}$ the \hhs relations (\ref{hh})
 imply the \hls relations (\ref{hl}) with $m_Q$ substituted by $m_{P_i}$.
Being more restrictive, eqs. (\ref{hh}) also provide the $q^2$ dependence
of the ratios of form factors and the $O(m_f/m_{P_i})$ corrections\cite{nous}.

An essential effect displayed by these  $O(m_f/m_{P_i})$ is that they soften
the
asymptotic scaling relation (\ref{hl}), i.e. they lead to a slower increase
(decrease) of $A_2$, $V$, $f_+$ ($A_1$) when the initial mass $m_{P_i}$
increases
at fixed $\vec q$.

Another welcome consequence of the Isgur-Wise relations (\ref{hh}) is that
they fix the ratios of form factors for the same quark masses and the same
transfer  $q^2$,  in such a way as to smoothen further
the inital mass dependence of the ratios when $q^2$
decreases. This is clearly illustrated at $q^2=0$ where all the form factors
are equal and their ratios, equal to 1, do not depend on the masses.

\subsection{Our Quark Model Inspired Ansatz.}
\label{sec-qmi}

We now formulate our model based on an extension of the \hh scaling relations
(\ref{hh}). Let us first assume that we are in a situation described in the
preceding section with two heavy quarks and $m_{i}\gg m_f\gg \Lambda$.
The form factors obey the \hls relations (\ref{hl}) with specific form factor
ratios and specific $O(m_f/m_i)$ corrections. To these, one
should also add the unknown $O(\Lambda/m_f)$ corrections to the heavy quark
symmetry.

Let us now consider the intermediate region where the final quark ceases to be
heavy. Our ignorance comes from the fact that the $O(\Lambda/m_f)$ corrections
become large and may totally modify the above mentioned specific relations. Our
hypothesis will be that it is not so, i.e. that using {\it some} of the
features
 of eq. (\ref{hh}) is indeed a good approximation. This hypothesis, although
admittedly
arbitrary, may be empirically justified to some extent as we shall see.
Theoretical
arguments in favor of the present Ansatz will come below and in section
3.6.

An unrestricted extension of Isgur Wise formulae (\ref{hh})
cannot describe quantitatively the form factors for a simple reason: the \dk
form factors at $q^2=0$, eq. (\ref{dexp}), are obviously not equal to each
other. To account for that we introduce some rescaling parameters
 ($r_+, r_V, r_1, r_2$), that we assume to be independent of the initial heavy
quark mass and of $q^2$.
 We therefore propose the following Ansatz that stays as close as possible
of the \hhs relations:

\[{  m_{P_i} + m_{P_f}   \over \sqrt{4 m_{P_i} m_{P_f}}}\left[ 1 - {q^2 \over
\left ( m_{P_i} +
m_{P_f} \right )^2}\right]\frac{ f_+(q^2)}{r_+} =
 {  m_{P_i} + m_{V_f}  \over \sqrt{4 m_{P_i} m_{V_f}}}
 \left[ 1 - {q^2 \over \left ( m_{P_i} +
m_{V_f} \right )^2} \right]\frac{V(q^2)}{r_V} =\]

\beq
=  {  m_{P_i} + m_{V_f}   \over \sqrt{4 m_{P_i} m_{V_f}}} \left[ 1 - {q^2 \over
\left ( m_{P_i} +
m_{V_f} \right )^2} \right]\frac{A_2(q^2)}{r_2} = {  m_{P_i} + m_{V_f}   \over
\sqrt{4 m_{P_i} m_{V_f}}} {A_1(q^2) \over r_1}  = \eta(\vec q, m_f)
\label{sp}\eeq
where $m_f$ is as usual the final meson mass: $m_{P_f}$ or $m_{V_f}$. In fact,
to conform with the asymptotic Isgur-Wise \hhs, the rescaling parameters $r_+,
r_V, r_2, r_1= 1+O(\Lambda/m_f)$, should depend on the final active quark mass
$m_{q_f}$ and tend to one when it goes to infinity, but this is irrelevant
here since the final quark will remain the $s$-quark all over this study.

This Ansatz has the wanted features of yielding  softened \hls relations and
a welcome $q^2$ dependence: $A_1(q^2)$ decreases as compared to the other
form factors when $q^2$ increases.

\subsection{Theoretical justifications of our Ansatz}
\label{sec-theor}
\vskip -0.35 cm

\paragraph {a) Quark model:}

How do we justify this Ansatz and in particular this ``rescaling'' procedure?
We are mainly motivated by the fact that the general structure of the
Isgur-Wise relations (\ref{hh})  also appears in the heavy to
light case in a quark model with weak binding treatment, the
Orsay Quark Model
(OQM)\cite{vieux}$^,$\cite{gif-91}$^{,25,26,5}$.
It gives the kinematical
 pole factor, differentiating
$f_+, A_2$ and  $V$ from $A_1$. It also displays the $O(m_f/m_i)$ corrections
predicted by the \hhs laws. On the other hand the quark model analysis leads to
expect two types of  $O(\Lambda/m_f)$ corrections:

i) Corrections  taking into account the finite mass of the spectator quark,
which are present in the weak binding treatment.

ii) Corrections to the weak binding limit, not included in the OQM.

In this model  the dominant correction to asymptotic scaling and the dominant
features of \qq dependence are represented by the Ansatz (\ref{sp}), while
additional corrections
are present but are small.

\paragraph {b) $B \to K^\ast \gamma$:}

An amusing example that exhibits the same trends as we advocate is
provided by the  $B \to K^\ast \gamma$ form factors. Defining the $T_i$ form
factors as follows,

\[
<K^\ast, k,\epsilon | \bar s \sigma^{\mu\nu} q_\nu \frac{1+\gamma_5}2 b|B,
p>=-
2 \epsilon_{\mu\nu\lambda\sigma}\epsilon^{\ast\nu}p^\lambda k^\sigma T_1(q^2)
 -i\bigg[\epsilon^\ast_\mu (m_B^2-m_{K^\ast}^2)-\]
\be \epsilon^\ast\cdot q (p+k)_\mu \bigg]
T_2(q^2)  -i \epsilon^\ast\cdot q\left[ q_\mu-\frac {q^2}{m_B^2-m_{K^\ast}^2}
(p+k)_\mu\right] T_3(q^2)\label{gamma}\ee
it is well known that, for $q^2=0$, using the identity
$\sigma_{\mu\nu}\gamma_5=\frac i 2
\epsilon_{\mu\nu\lambda\sigma}\sigma^{\lambda\sigma}$, one obtains the exact
relation:

\be T_1(0)=T_2(0).\label{egalite}\ee

It has also been shown\cite{iwbsg} that

\[ T_1(q^2)= \sqrt{m_Q}\left(1 + O\left(\frac \Lambda{m_Q}\right)+
O\left(\frac {|\vec q|}{m_Q}\right)\right),\]
\be T_2(q^2)= \frac 1 {\sqrt{m_Q}}\left(1 + O\left(\frac \Lambda{m_Q}\right)+
O\left(\frac {|\vec q|}{m_Q}\right)\right).\label{hlbsg}\ee

In the \hh case one may also show that

\be{ \sqrt{4 m_{P_i} m_{V_f}} \over m_{P_i} + m_{V_f} } T_1(q^2)=
{\sqrt{4 m_{P_i} m_{V_f}} \over m_{P_i} + m_{V_f} } \frac {T_2(q^2)}
{1 - {q^2 \over \left ( m_{P_i} +
m_{V_f} \right )^2}}=\frac 1 2\xi(v\cdot v'),\label{hhbsg}\ee
which is of course fully compatible with the relation (\ref{egalite}).

The novelty here is that {\it the relation (\ref{egalite}) remains exact when
the final quark becomes light}. Since the scaling behaviours of the $T_1$ and
$T_2$ differ in the vicinity of $q^2_{max}$, eq. (\ref{hlbsg}), the equality
(\ref{egalite}) is {\it a clear indication that the \qq behaviour of both form
factors differs sensibly.}
 For example a pole dominance hypothesis for both form factors is totally
excluded by these relations. Furthermore, an extension of
relation (\ref{hhbsg}) to the \hl domain, as we have suggested in section
3.5, would directly comply with both relations (\ref{egalite}) and
(\ref{hlbsg}). Lattice calculations may indicate\cite{glasgow} a rather flat
$q^2$
dependence of $T_2(q^2)$ except one lattice group\cite{soni} who finds, on the
contrary, a
 pole-like $T_2(q^2)$. The flat behaviour leads to a $B\to K^\ast\gamma$
branching ratio
in agreement with experiment, as expected since the long distance
contributions,
which are overlooked by the lattice,
have been estimated\cite{golowich} to be small.

 Finally, let us  insist that this is by no means a proof of our
Ansatz, it is simply a hint that it points towards the right direction.

\paragraph {c) Matrix elements:}

Some light can be shed on our Ansatz (\ref{sp}), as far as the mass dependence
is concerned, by noting that it amounts to assume that {\it the matrix elements
satisfy an uncorrected asymptotic scaling}. To illustrate this point let us
consider a final vector meson $V_f$ with a polarization $\epsilon^T$ orthogonal
to the initial and final meson momenta.

{}From eqs. (\ref{definite}) and (\ref{sp}) the matrix elements scale as
follows:

\[ \frac{<V_f, \epsilon^T, \vec q |A_\mu|P_i>}{\sqrt{4m_B m_{V_f}}} =r_1
\,\eta(\vec q, m_{V_f}) \epsilon_\mu,\]

\be \frac{<V_f, \epsilon^T, \vec q |V_\mu|P_i>}{\sqrt{4m_B m_{V_f}}}= i r_V \,
\frac{\eta(\vec q, m_{V_f})} {1+v^0_f} \,{\left(\vec v_f \times \vec
\epsilon\,^T\right)_\mu},\label{matel}\ee
where $v_f^\mu=p_f^\mu/m_f$  has been used as well as the relation:
\be 1- \frac{q^2}{(m_{P_i}+m_f)^2}= 2\frac{ m_{P_i}(m_f+E_f)}{(m_{P_i}+m_f)^2}.
\label{energie}\ee.

In this example it is clear that the matrix elements scale exactly like
$\sqrt{m_{P_i} m_f}$, which is their asymptotic \hls behaviour. Our claim in
favor of the softened scaling, eq. (\ref{sp}), is equivalent to the statement
that
the matrix element asymptotic scaling laws are not corrected at non-asymptotic
masses.
In other words {\it our softened scaling Ansatz is equivalent to a precocious
asymptotic scaling of the matrix elements}.

\paragraph {d) QCD sum rules, lattice calculations:}

We have argued in\cite{nous} that QCD sum rules qualitatively
favor\cite{dosch}$^,$\cite{ball}$^,$\cite{ali}$^{,33,34}$
the \qq dependence of the form factor ratios as depicted in eq. (\ref{sp}),
i.e. they generally show an increase of the ratios $A_2/A_1, V/A_1, f_+/A_1$,
with \qq not very different from the increase due to the kinematical pole
$1/(1-q^2/(m_{P_i}+m_f)^2)$.

Lattice calculations\cite{abada} on their side, favor a softened \hls, as a
function of the heavy masses, for the leptonic decay constant $F_P$ and for the
form factors except $A_2$. However, if this result seems strongly established
for the leptonic decay constants, the $m_Q$ dependence of the form factors, and
particularly $A_2$, are not yet known with enough precision  to be conclusive.

\subsection{Confronting the form factors to experimental ratios.}
\label{sec-confr}
\vskip -0.35 cm

In ref.\cite{nous} we have performed a simultaneous $\chi^2$ fit of the
\bpsi and \dk data, combining them with the help of different Ans\"atze.
The latter Ans\"atze are devised to compare
 the most commonly used ``natural'' assumptions with our favored eq.
(\ref{sp}).
 For instance they correspond to
uncorrected asymptotic scaling of the form factors
and/or $q^2$ independent form factor ratios\footnote{Universal
pole dominance, a very popular assumption, implies approximately constant
form factor ratios.}.

For every Ansatz we have performed a least $\chi^2$ fit\cite{nous}. If we
only fit the final polarization, i.e. the ratio $R_L$ defined in eq.
(\ref{rl}),
 in \bpsi with the \dk data, our Ansatz (\ref{sp}) gives
the smallest $\chi^2$. This was expected since (\ref{sp}) reduces
$A_2^{sb}/A_1^{sb}$ and
therefore improves $R_L$, see eq. (\ref{rlform}). However this improvement  is
insufficient,
since the best $\chi^2$ thus obtained is
$\sim 2 (\sim 4)$ per degree of freedom when $R_L$ is taken from CDF (CLEO II),
i.e.
 a 2-3 $\sigma$ discrepancy.
When we fit both $R_L$ and $R$, eq. (\ref{rstar}), with \dk data, the Ansatz
(\ref{sp})
gives a least $\chi^2$ of $\sim 3 (\sim 4)$ per d.o.f, not better than that the
alternative
Ansatz which incorporates also softened scaling, but $q^2$ independent form
factor ratios.
The reason is that all our trial Ans\"atze assume $A_2/A_1=f_+/A_1$, and, as
argued
after eq. (\ref{beau}), the $R_L$ data demand a smaller $A_2/A_1$ while $R$
data demand a
larger $f_+/A_1$. The $\chi^2$ fit tries a compromise between these opposite
trends.
Only by relaxing the constraint $A_2/A_1=f_+/A_1$ can this be cured, as done in
ref\cite{cheng}.

To summarize:

\begin{itemize}

\item As anticipited, the experimental comparison between \bpsi and \dk favors
a soft \hl
scaling.

\item The best \qq dependence cannot be selected from this analysis alone.
The separated phenomenological study of the $K^\ast$ final states,
 as well as several theoretical considerations, tend to
favor the existence of the ``kinematical pole''.
But the consideration of the \gs ratio tends to wash out this conclusion.

\item There remains a difficulty to reconcile experimental results in \bpsi
and \dk when taking CDF results for $R_L$ ($\chi^2/dof \simeq 3$), which
worsens
when using CLEO or ARGUS values for $R_L$. There seems to be also a
particular difficulty to fit simultaneously $R$ and $R_L$. Only fragile
indications of possible ways out of these difficulties are known
today\cite{nous}.

\end{itemize}

\section{Conclusions.}
\label{sec-conc}

The first requirement for any model is to fulfill the \hls
relations. This is not the case\cite{nous} for the most popular BSW I and
BSW II models, notwithstanding their relatively good empirical successes.
ISGW does not fulfill the \hhs relations and fails very badly for the
polarization.

All the approaches we have considered in this paper encounter difficulties in
accounting for the \bpsi data, particularly with the large \glgt (CLEO and
ARGUS data). At present it seems safer to keep  open three possibilities to get
out of this problem.
\begin{itemize}

\item Experiment may not have yet delivered its ultimate word, as the variation
between different experiments seem to indicate, and it might evolve towards
data easier to account for.

\item Although we did not discuss the factorization assumption, it should be
kept in mind that it rests on no theoretical ground for color suppressed decay
channels, as is the case for \bpsi. Carlson and Milana\cite{milana} find a
 significant correction to factorization within their perturbative QCD
inspired model. Gourdin, Keum and Pham\cite{keum}, propose
to test  factorization in $B\to \eta_c K^{(\ast)}$.

\item Finally, models may be wrong. This will now be discussed in more details.
\end{itemize}

Our analysis has allowed to extricate from data some general trends, namely
``softened'' scaling, a sensibly different \qq behaviour of $A_1$ versus $A_2,
V, f_+$, and $A_1$ slowly varying with \qq. Ans\"atze that take these
indications as a guide, obtain better values for \glgt, and a more reasonable
$a_2$ (see ref.\cite{nous}), although
there remains a general tendency to underestimate \glgt with respect to present
data.
Let us now comment on these general trends.

Data definitely exclude ``hard scaling'' i.e. the strict application of
asymptotic \hls formulae in the finite mass domain. We have proposed a
``softened'' Ansatz which is based on an extension of \hhs relations down to
the light final meson case, with some rescaling. In fact this is equivalent to
assuming a precocious scaling for the axial and vector current {\it matrix
elements}. Consequently, the ratio $A_2/A_1$ does not increase too fast with
the heavy mass.

There are indications from lattice calculations, form Quark Model, and to some
degree from phenomenology, that $V$ should undergo an even softer
scaling\cite{nous}.

Another consequence of the above Ansatz, as well as of the Orsay Quark Model is
that $A_2/A_1, V/A_1$ and $f_+/A_1$ should have a pole like behaviour in \qq,
leading to an increase with \qq. This improves the agreement with \bpsi data,
and seems to be corroborated by QCD Sum Rules calculations.

$D\to Kl\nu$ experiments seem to show a pole like behaviour for $f_+(q^2)$.
Combined with our preceding Ansatz for the ratios, this implies an
approximately constant $A_1(q^2)$. This particular \qq behaviour is
corroborated
by the Orsay Quark Model, while QCD Sum Rules give a \qq dependence of $A_1$
that
never increases very fast, although different detailed shapes are proposed.
Lattice calculations, within large errors, might give the same indication.

\section{Acknowledgements.}
We acknowledge  Nathan Isgur,  Stephan Narison and Daryl Scora for interesting
comments and useful informations.
We are particularly indebted to Patricia Ball and to the European Lattice
Collaboration for providing us with their results concerning semi-leptonic form
factors, to  Asmaa Abada, Philippe Boucaud and Jean Pierre Leroy for
enlightening discussions and to Belen Gavela for several discussions
and a critical reading of this manuscript.
This work was supported in part by the CEC Science Project SC1-CT91-0729 and by
the Human Capital
and Mobility Programme, contract CHRX-CT93-0132.


\begin{thebibliography}{9}
\bibitem{oddone}P. Oddone, these proceedings.
\bibitem{alek18} R. Aleksan, A. Le Yaouanc, L. Oliver, O. Pene and J.-C.
Raynal, Phys. Lett. \underbar{B316}, 567 (1993).
\bibitem{alek03} R. Aleksan, A. Le Yaouanc, L. Oliver, O. Pene and J.-C.
Raynal, LPTHE-ORSAY-94-03, Jun 1994, hep-ph:9407406.
\bibitem{beauty-94} R. Aleksan et al., Beauty 94 workshop, April 1994, Mont
Saint Michel, France (to be published in Nucl. Inst. and meth. A), LPTHE Orsay
94/53 Bulletin Board: hep-ph 9406334.

\bibitem{nous}R. Aleksan, A. Le Yaouanc, L. Oliver, O. P\`ene and J.-C. Raynal,
 DAPNIA/SPP 94-24, LPTHE-Orsay 94/15,
Bulletin Board: hep-ph 9408215.
\bibitem{gourdin} M. Gourdin, A.N. Kamal and X.Y. Pham, PAR/LPTHE/94-19.
\bibitem{svz}M.A. Shifman, A.I. Vainshtein and V.I. Zakharov, Zh. Ekxp. Teor.
Fiz. \underbar{72}, 1275 (1977) [Sov. Phys. JETP \underbar{45}, 670 1977)].
\bibitem{bsw1} M. Wirbel, B. Stech and M. Bauer, Z. Phys. \underbar{C29}, 637
(1985); \underbar{C34}, 103 (1987).
\bibitem{bsw2} M. Neubert, V. Rieckert, B. Stech and Q.P. Xu, in
\underline{Heavy Flavours}. eds. A.J. Buras and M. Lindner, World Scientific,
Singapore (1992).
\bibitem{cleo-94} S. Henderson, CLEO collaboration, Talk given at XXIXth
Rencontre de Moriond, ``QCD and High Energy Hadronic Interactions'', March
1994, Ed. Fronti\`eres; D. Payne, Talk delivered at ``Beauty 94'', Mont
Saint-Michel, Avril 1994.
\bibitem{cleo2} CLEO II, M. Danilov, Talk given at the ECFA Working group on B
Physics, DESY (1992).
\bibitem{cdf} CDF, G. Apollinari and K. Goulianos, private communication.
\bibitem{browder}T.E. Browder, K. Honscheid and S. Playfer, CLEO Report CLNS
93/1261, 1994, to appear \underbar {B decays}, 2nd. Edition, Ed. S. Stone,
World Scientific, Singapore.
\bibitem{isgw}N. Isgur, D. Scora, B. Grinstein and M.B. Wise, \prd{D39}
(1989) 799; N. Isgur and D. Scora \prd{D40} (1989) 1491.
\bibitem{ball} P. Ball, Phys. Rev. \underline{D48} (1993) 3190A.
\bibitem{ballp} P. Ball, private communication.
\bibitem{abada} A. Abada et al. Nucl. Phys. \underbar{B416}, 675 (1994).
\bibitem{witherell}M. S. Witherell, Invited talk at International Symposium on
Lepton and Photon Interactions at High Energies, Cornell, Ithaca, N.Y.,
UCSB-HEP-93-06. The  used
data are taken among others from: P.L. Frabetti et al. (E-687) Phys. Lett.
\underbar{B307}, 262 (1993); J.C. Anjos et al. (E-691) Phys. Rev. Lett.
\underbar{62}, 722 (1989); \underbar{65}, 2630 (1990); K. Kodama et al. (E-653)
Phys. Lett. \underbar{B274}, 246 (1992).
\bibitem{isgurhl} N. Isgur and M.B. Wise Phys. Rev. \underbar{D42}, 2388
(1990).
\bibitem {cheng} H-Y Cheng and B. Tseng, IP-ASTP-21-94, hep-ph:  9409408.
\bibitem{isgur} N. Isgur and M.B. Wise, Phys. Lett. \underbar{232B}, 113 (1989)
and \underbar{237B}, 527 (1990).
\bibitem{nr} M. Neubert and V. Rieckert, Nucl. Phys. \underbar{B382}, 97
(1992).
\bibitem{vieux} P. Andreadis et al. Ann. of Phys. \underbar{88} (1974) 242;
A. Amer et al., Phys. Lett. \underbar{81B}  (1979) 48; M.B. Gavela,
Phys. Lett. \underbar{83B}  (1979) 367; M.B. Gavela, Doctoral thesis
univ. Paris-sud (LPTHE), July 1779.
\bibitem{gif-91} A. Le Yaouanc et al., Gif lectures 1991 (Electroweak
properties of heavy quarks), tome 1, p89.
\bibitem{marbella-93} A. Le Yaouanc and O. P\`ene, Third workshop on the
Tau-Charm Factory, 1-6 June 1993, Marbella, Spain, Bulletin Board: hep-ph
9309230.
\bibitem{oqm} A. Le Yaouanc et al. in preparation.
\bibitem{iwbsg} N. Isgur and M.B. Wise Phys. Rev \underbar{D42} (1990) 2388 .
\bibitem{glasgow}  As. Abada, Proceedings  of the XXVII Int.
 Conf. on High Energy Phys., Glasgow, July 1994, hep-ph: 9409338;
As Abada et al. APE collaboration, LPTHE Orsay-94/88 and ROME prep. 94/1056.
\bibitem{soni} A. Soni Proceedings  of the XXVII Int.
 Conf. on High Energy Phys., Glasgow, July 1994, hep-ph: 9410007.
\bibitem{golowich} E. Golowich and S. Pakvasa, UMHEP-411, UH-511-800-94,
hep-ph:  9408370.
\bibitem{dosch}P.Ball, V.Braun and H.G.Dosch, Phys. Rev. D44 (1991) 3567.
\bibitem{ali} A.Ali, V.Braun and H.Simma, CERN-TH. 7118/93.
\bibitem{belyaev} V.M. Belyaev, A. Khodjamirian and R. R\"uckl, Z. Phys.
\underbar{C60}, 349 (1993).
\bibitem{narison}S. Narison, CERN-TH.7237/94.
\bibitem{keum} M. Gourdin, Y.Y. Keum and X.Y. Pham , PAR-LPTHE-94-32, Bulletin
Board:
 hep-ph@xxx.lanl.gov - 9409221.
 \bibitem{milana} C. E. Carlson and J. Milana , WM-94-110, Bulletin Board:
hep-ph@xxx.lanl.gov - 9409261.


\end{thebibliography}
\end{document}